\documentstyle[11pt,newpasp,twoside,epsf]{article}
\def\edcomment#1{\iffalse\marginpar{\raggedright\sl#1\/}\else\relax\fi}
\reversemarginpar

\begin{document}
\title{PSR B0656+14: Combined Optical, X-ray \& EUV Studies}
 \author{Aaron Golden, Andy Shearer}
\affil{The National University of Ireland,
Galway, Newcastle Road, Galway, Ireland}
\author{Jerry Edelstein}
\affil{Space Sciences Laboratory, UC Berkeley, USA}

\begin{abstract}

PSR B0656+14's high energy emission is consistent 
with that of combined magnetospheric and thermal (surface
\& polar cap) emission. 
Uncertainties with the radio-derived distance and
X-ray instrumentation sensitivities complicate a definitive
thermal characterisation however.
A re-analysis of combined ROSAT/EUVE archival data
in conjunction with integrated \& phase-resolved optical
photometry is shown to constrain this characterisation.

\end{abstract}

\section{Introduction}

Considerable uncertainty remains regarding the 
fundamental thermal parameters ($T,N_{H} \& R/d$) for PSR B0656+14.
Radio derived DM estimates (790 $\pm$ 190 pc) disagree with
the best $N_{H}$ model fits (250 -- 550 pc). Reported calibration uncertainties
associated with the low energy channels of the $ROSAT$ PSPC compromise
the latter - although agreement between other $ROSAT$ PSPC 
\& observed $EUVE$ fluxes 
obtained via a correction (e.g. for RX J185635-3754, Walter \& An, 1998).
We outline the results of such a correction to the existing PSPC datasets archived
for PSR B0656+14 via substitution of the low energy channels with 
measured $EUVE$ fluxes, and by incorporating independently derived
constraints to the Rayleigh-Jeans tail in the optical, discuss
the implications for the neutron star's thermal parameters.

\section{Technical \& Analytical Overview}

Optimum thermal
fits for $T_{soft}, T_{hard}, N_{H}, R/d$ were obtained for the archived $ROSAT$ PSPC data
alone and the PSPC data with the suspect low energy channels substituted with
the archival normalised EUVE flux. This substitution results in a significant change 
in solution space, as shown in Figure 1 (Edelstein et al. 1999). Based on integrated
optical photometry, Pavlov et al. (1997) fitted a two component
nonthermal/thermal model, the thermal fit defined by a parameter
$G$ $\equiv$ $T_{10^{6}\rm K} {(R_{10\rm km} / d_{500\rm pc})}^2$
where $G$ = [1 -- 7] (see Figure 1).
\begin{figure}
\plottwo{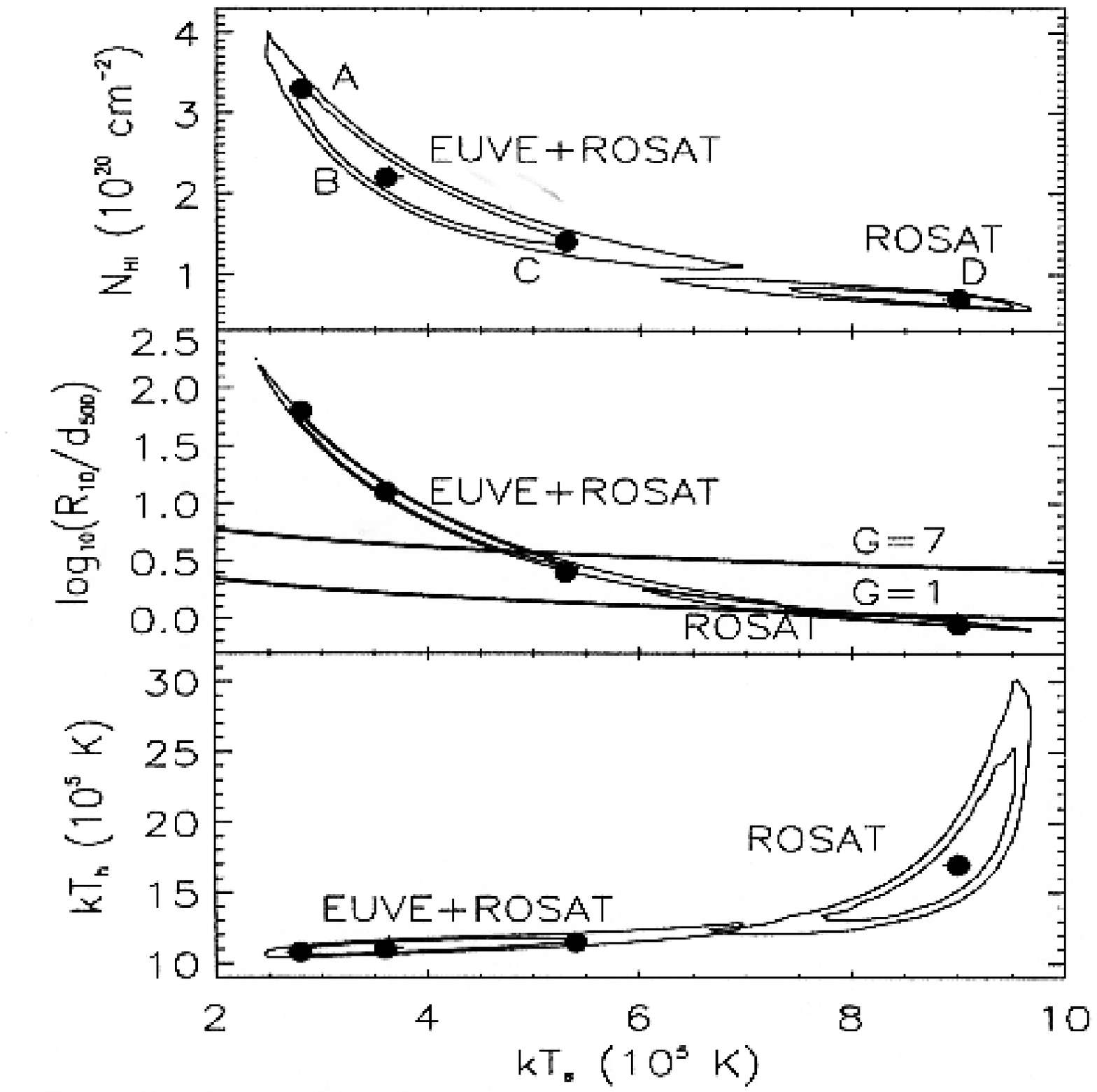}{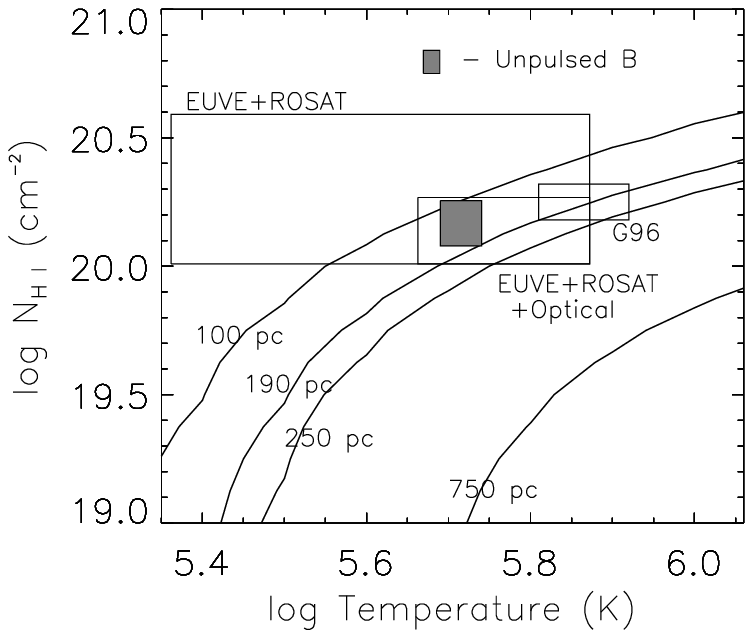}
\caption{\scriptsize{{\bf Left:} Model Fits to $T_{h}, N_{H},R/d$ \rm {vs.} $T_{s}$ at 
the 90\% and 99\% confidence interval for
ROSAT and EUVE+ROSAT datasets. The range of best-fit $G$ parameters  
are indicated. {\bf Right:} ($N_{H},T_{s}$) space constrained by EUVE \& ROSAT, and incorporating the
optical constraints to $G$. Curves are locii of constant $(N_{H},T_{s})$ based on the observed
EUVE count rate. Previous solution of Greiveldinger et al. (1986) is marked.}}
\end{figure}
A 1$\sigma$ upper limit on the unpulsed component from the optical
$B$ band light curve of Shearer et al. (1997) limits $G$ $\le$ 4.4, 4.8 and 5.2,
based on various optical extinction models to the pulsar (Golden, 1999). These optical
results yields tighter constraints on parameter space,
as can be seen.  

\section{Discussion \& Conclusions}

Combining the EUVE \& ROSAT datasets in this way yields new solutions
in parameter space that are further constrained {\it independently} via
recent optical work. Assuming a simple blackbody form then  
$T_{surface}$ $\geq$ 5.0$\times 10^{5}$ K and for
the $N_{H}$-derived distances of [250 -- 280] pc, 
$R_{\infty}$ $\le$ [17.7 -- 14.7] km.
Using the estimate of $R_{\infty}$ $\sim$ $9.5 ^{+3.5}_{-2.0}$ km 
for Geminga as a working upper limit (Golden \& Shearer, 1999)
places PSR B0656+14 at a distance of no less than 
$d$ = $152 ^{+55}_{-32}$. This suggests the possibility
of parallax observations to independently derive $d$, with
immediate implications for the $R$ parameter, and consequently
models of the condensed matter equation of state.


\begin{references}
\reference{Edelstein, J., Seon, K.-I., Golden, A., \& Kwok-in, K., 1999, sub. ApJ.}
\reference{Golden, A., 1999, Ph.D. Thesis, {\it National University of Ireland, Galway}.}
\reference{Golden, A., \& Shearer, A., 1999, A\&A, 342, L5.}
\reference{Greiveldinger, C., Camerini, U., Fry, U., Markwardt, C.B., Ogelman, H., et al., 1986, ApJ, 465, 35.}
\reference{Pavlov, G.G., Welty, A.D., \& Cordova, F.A., 1997, ApJ, 489, L75.}
\reference{Shearer, A., Redfern, R.M., et al., 1997, ApJ, 487, L181.}
\reference{Walter, F.M., \& An, P., 1998, AAS, 192, 82.07.}
\end{references}
\end{document}